\documentclass[aps,pra,twocolumn,amsmath,amssymb,nofootinbib,showpacs,superscriptaddress]{revtex4-1}
\usepackage[english]{babel}
\usepackage{latexsym}
\usepackage{graphics}
\usepackage{graphicx}
\usepackage{epsfig}
\usepackage{color}
\usepackage{bm}
\usepackage{amsmath}
\usepackage{amssymb}
\usepackage{amsthm}
\usepackage{dcolumn}
\usepackage{bm}
\usepackage{float}
\usepackage{hyperref}
\usepackage{color}
\usepackage{epstopdf}
\usepackage{braket}
\usepackage[svgnames]{xcolor}
\usepackage{enumerate}
\hypersetup{hidelinks,colorlinks=true,allcolors=DarkBlue}

\usepackage{lineno}

\begin{document}
	
	\preprint{APS/123-QED}
	
	\title{Protecting quantum systems from decoherence with unitary operations}
	
	\author{E.O. Kiktenko}
	\affiliation{Russian Quantum Center, Skolkovo, Moscow 143025, Russia}
	\affiliation{Moscow Institute of Physics and Technology, Dolgoprudny, Moscow Region 141700, Russia}
	\affiliation{Department of Mathematical Methods for Quantum Technologies, Steklov Mathematical Institute of Russian Academy of Sciences, Moscow 119991, Russia}
	
	\author{A.S. Mastiukova}
	\affiliation{Russian Quantum Center, Skolkovo, Moscow 143025, Russia}
	\affiliation{Moscow Institute of Physics and Technology, Dolgoprudny, Moscow Region 141700, Russia}
	
	\author{A.K. Fedorov}
	\affiliation{Russian Quantum Center, Skolkovo, Moscow 143025, Russia}
	\affiliation{Moscow Institute of Physics and Technology, Dolgoprudny, Moscow Region 141700, Russia}
	
	\date{\today}
	\begin{abstract}
		Decoherence is a fundamental obstacle to the implementation of large-scale and low-noise quantum information processing devices.
		In this work, we suggest an approach for suppressing errors by employing pre-processing and post-processing unitary operations, which precede and follow the action of a decoherence channel.
		In contrast to quantum error correction and measurement-based methods, the suggested approach relies on specifically designed unitary operators for a particular state without the need in ancillary qubits or post-selection procedures.
		We consider the case of decoherence channels acting on a single qubit belonging to a many-qubit state.
		Pre-processing and post-processing operators can be either individual, that is acting on the qubit effected by the decoherence channel only, or collective, that is acting on the whole multi-qubit state.
		We give a classification of possible strategies for the protection scheme, analyze them, and derive expressions for the optimal unitary operators providing the maximal value of the fidelity regarding initial and final states.
		Specifically, we demonstrate the equivalence of the schemes where one of the unitary operations is individual while the other is collective.
		We then consider the realization of our approach for the basic decoherence models, which include single-qubit depolarizing, dephasing, and amplitude damping channels.
		We also demonstrate that the decoherence robustness of multi-qubit states for these decoherence models is determined by the entropy of the reduced state of the qubit undergoing the decoherence channel.
	\end{abstract}
	
	\maketitle
	
	\section{Introduction}
	
	A concept of quantum internet relies on the ability of quantum computers to process and exchange an essentially arbitrary number of quantum states of any structures~\cite{Kimble2008}.
	Although key building blocks both for quantum computing devices and quantum communication systems have been developed~\cite{Ladd2010,Lukin2016,Jiang2016,Martinis2017,Wehner2018}, 
	the current challenge is to scale such devices with respect to the number of qubits inside quantum computers and the distance of quantum state transfer~\cite{Wehner2018}.
	A major barrier is protecting quantum systems from decoherence, which is the main source of errors in quantum information processing devices~\cite{Ekert1996,Duan1998,Chuang1998,Buchleitner2009,Breuer2002,Zurek2003}.
	An ultimate solution for eliminating the decoherence effect on quantum systems 
	would be the use of quantum error correction~\cite{Kitaev1997,Bennett1996,Steane1997,Laflamme2000,Gottesman2010,Jiang2009,Munro2010,Fowler2010,Jiang2014,Jiang2017}.
	However, existing methods are hardly implementable since they typically require additional resources, such as ancillary qubits, 
	or they are demanding from the viewpoint of code sizes~\cite{Gottesman2010,Jiang2009,Munro2010,Fowler2010,Jiang2014,Jiang2017}.
	
	Here, instead of considering quantum error correction methods, we stress on possible methods for error suppression that would reduce, but not eliminate the effect of decoherence in transferring quantum states.
	This problem has been studied in recent decades in various aspects. 
	Particular schemes include but not limited to the use of non-Gaussian states~\cite{Filippov2014}, 
	controlling information asymmetry in multi-qubit systems~\cite{Zyczkowski2001,Kiktenko2012,Korotaev2012,Kiktenko2014},
	as well as employing quantum Zeno effect~\cite{Gurvitz2003,Facchi2004,Maniscalco2008}, specific measurements~\cite{Korotkov2010,Kim2012,Roszak2015,Basit2017} and many other methods control.
	These methods are useful under certain conditions, however, they encounter some difficulties for the experimental realization. 
	A particular task is improving the level of entanglement in a distributed quantum state after its degradation due to losses in communication lines, which can be solved efficiently with the use of quantum catalysis~\cite{Lvovsky2015}. 
	Unfortunately, the entanglement level increases at the cost of employing post-selection.

	\begin{figure}[t]
		\begin{centering}
			\includegraphics[width=1\columnwidth]{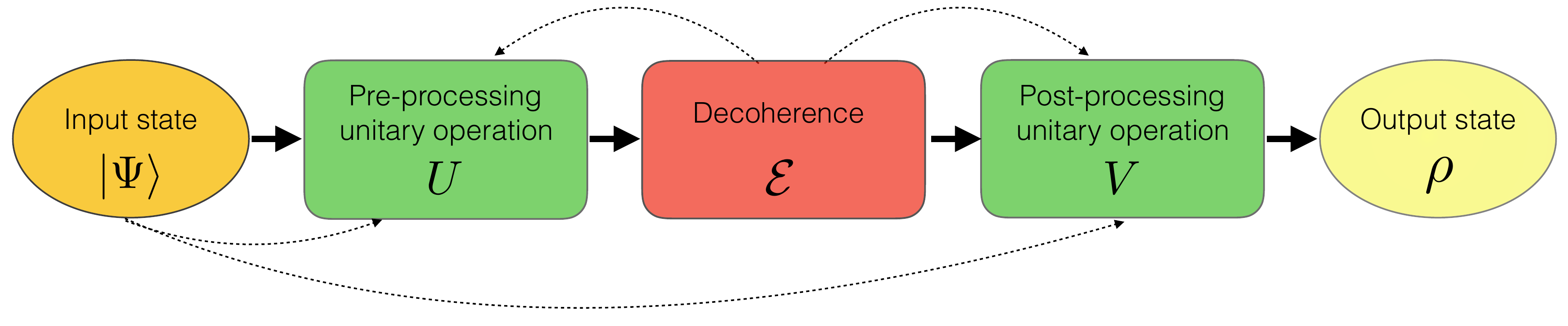}
		\end{centering}
		\vskip -8mm
		\caption{Error suppression scheme based on pre-processing and post-processing unitary operations, which are designed specifically for the input state and decoherence channel in order to maximize the fidelity of the output state.}
		\label{fig:genscheme}
	\end{figure}
	
	Here we propose a technique for protecting quantum states from decoherence, which is free from using ancillary information carriers or post-selection procedures (see Fig.~\ref{fig:genscheme}).
	First, we use a {\it pre-processing} procedure for preparing a given known quantum state of the system in a specific form.
	Next, we use a {\it post-processing} operation, which follows the action of a decoherence channel.
	These operations can be implemented as unitary operators, and their particular form can be efficiently constructed based on prior knowledge of the state under the protection and decoherence channel, i.e. the method is \emph{state-dependent}.
	We study the case of local decoherence channels, which act on a particular qubit of the $n$-qubit input state.
	The pre-processing and post-processing unitary operators are considered to be either individual, that is acting on the same qubit, or collective, that is acting on the whole $n$-qubit state.
	As a model for decoherence processes, we consider single-qubit depolarizing, dephasing, and amplitude damping channels.
	In our consideration, the main studied characteristic is the fidelity regarding input and output states.
	We present an analysis of possible strategies for the protection scheme and derive expressions for the optimal unitary operators providing the maximal value of the fidelity regarding initial and final states.
	Specifically, we show the equivalence of the schemes where one of the unitary operations is individual while the other is collective.
	
	We also consider relations between the form of the reduced state of the qubit undergoing an individual decoherence process and losses of the fidelity of the whole $n$-qubit state.
	We demonstrate that in the case of performing pre- and post-processing this kind of the robustness to decoherence is determined by the linear entropy the reduced state.
	This feature appears to be crucial in the case where one can select a part of the whole quantum state, which is affected by the decoherence.
	It is important to note that the suggested scheme is able in principle to supplement 
	existing error correction and error suppression techniques~\cite{Kitaev1997,Bennett1996,Steane1997,Laflamme2000,Gottesman2010,Jiang2009,Munro2010,Fowler2010,Jiang2014,Jiang2017,Fedichkin2000,Ozhigov2003}. 
	
	Our paper is organized as follows.
	In Sec.~\ref{sec:schemes}, we present a general framework for a method of protecting quantum states from decoherence with unitary operations.
	In Sec.~\ref{sec:models}, we apply the presented method for the case of acting basic single-qubit decoherence channels, which include depolarizing, dephasing, and amplitude damping channels.
	In Sec.~\ref{sec:conclusion}, we discuss the obtained results and conclude.
	
	\section{Protecting states from decoherence with unitary operations}\label{sec:schemes}
	
	Consider a $n$-qubit quantum system, which consists of $n$ subsystems with indices $1,\ldots,n$.
	Let the whole system be initialized in a pure state written in the following form:
	\begin{equation}\label{eq:instate}
	\ket{\Psi}_{1,\ldots,n} = \sum_{{\bf x}\in\{0,1\}^n} c_{\bf x} \ket{\bf x}_{1,\ldots, n},
	\end{equation}
	where complex coefficients $\{c_{\bf x}\}$ obey a standard normalization condition $\sum_{{\bf x}\in\{0,1\}^n} |c_{\bf x}|^2 = 1$.
	Then let a particular $\kappa$th qubit ($\kappa$$\in\{1,\ldots,n\}$) undergo a decoherence process described by a completely positive trace-preserving (CPTP) map $\mathcal{E}$, which we refer as a \emph{decoherence channel}.
	The initial state~\eqref{eq:instate} then turns into a new (generally, mixed) state as follows:
	\begin{equation}
	\rho_{1,\ldots,n}^{(\kappa)}={\rm Id}_{\{1,\ldots,n\}/\{\kappa\}}\otimes \mathcal{E}_{\kappa} [\ket{\Psi}_{1,\ldots,n}\bra{\Psi}],
	\end{equation}
	where ${\rm Id}_{\{1,\ldots,n\}/\{i\}}$ denotes the identical channel acting on all qubits except $\kappa$th and $\mathcal{E}_\kappa$ is a decoherence channel acting on $\kappa$th qubit.
	As the main target characteristic, we consider the fidelity regarding input and output states, which is given by the following expression:  
	\begin{equation}
	F := \bra{\Psi}_{1,\ldots,n} \rho_{1,\ldots,n}^{(\kappa)} \ket{\Psi}_{1,\ldots,n}.
	\end{equation}
	In the general case, the state $\rho_{1,\ldots,n}^{(\kappa)}$ has less than one fidelity with respect to $\ket{\Psi}_{1,\ldots,n}$. 
	
	The main goal of our work is to develop a method for improving the fidelity of the state after decoherence by means of employing unitary operations acting on qubits.
	We study using two types of unitary operations: 
	\begin{itemize}
		\item \emph{individual} operations acting on $\kappa$th qubit only; 
		\item \emph{collective} acting on the whole $n$-qubit system.
	\end{itemize}
	Specifically, we consider employing two unitary operations just before and after an impact of the decoherence channel $\mathcal{E}$.
	We refer to these operations as \emph{pre-processing} and \emph{post-processing} unitary operators.
	Since each of the processing operators can be either individual or collective, we obtain four possible strategies: 
	\begin{enumerate}
		\item `both individual' scheme: both operations are individual; 
		\item `individual-then-collective' scheme: pre-processing operation is individual, while the post-processing operation is collective; 
		\item `collective-then-individual' scheme: pre-processing operation is collective, while the post-processing operation is individual; 
		\item `both collective' scheme: scheme with two collective operations.
	\end{enumerate}
	
	We denote $\kappa$th qubit as $Q$, while the subsystem of all the rest qubits $\{1,\ldots,n\}/\{q_i\}$ as $R$. 
	Using the Schmidt decomposition, the initial state~\eqref{eq:instate} can be written in the following form:
	\begin{equation}
	\ket{\Psi}_{QR} = \sum_{i=0,1} \sqrt{\lambda_i} \ket{\psi_i}_Q \otimes \ket{\zeta_i}_R,
	\end{equation}
	where a pair of states $\{\ket{\psi_0},\ket{\psi_1}\}$ forms the orthonormal basis in the Hilbert space of $Q$, $\{\ket{\zeta_0},\ket{\zeta_1}\}$ are two orthogonal normalized vectors in the space of $R$, 
	$\lambda_0$ and $\lambda_1$ are non-negative real numbers such that $\lambda_0+\lambda_1=1$ and $\lambda_0\geq\lambda_1$, which always can be achieved by an appropriate choice of $\{\ket{\psi_i}\}$ and $\{\ket{\zeta_i}\}$.
	To characterize the information properties of the considered quantum system, we also employ linear entropy,
	\begin{equation}
	S_{\rm lin}(\rho) := 1-{\rm tr}(\rho^2),
	\end{equation}
	von Neumann entropy,
	\begin{equation}
	S_{\rm vN}(\rho) := -{\rm tr}(\rho\log\rho),
	\end{equation}
	and min-entropy,
	\begin{equation}
	S_{\rm min}(\rho):= -\log\Lambda_{\max}(\rho),
	\end{equation}
	where $\log$ stands for base-2 logarithm, $\rho$ is a density matrix of arbitrary dimension, and $\Lambda_{\max}(\rho)$ is operator giving the largest eigenvalue of $\rho$.
	We note that von Neumann entropy and min-entropy are special cases of the R\'enyi entropy:
	\begin{equation}
	S_\alpha(\rho):=\frac{1}{1-\alpha}\log{\rm tr}(\rho^\alpha)
	\end{equation} 
	for $\alpha\rightarrow 1$ and $\alpha\rightarrow+\infty$, correspondingly.
	
	\begin{figure*}
		\includegraphics[width=0.7\linewidth]{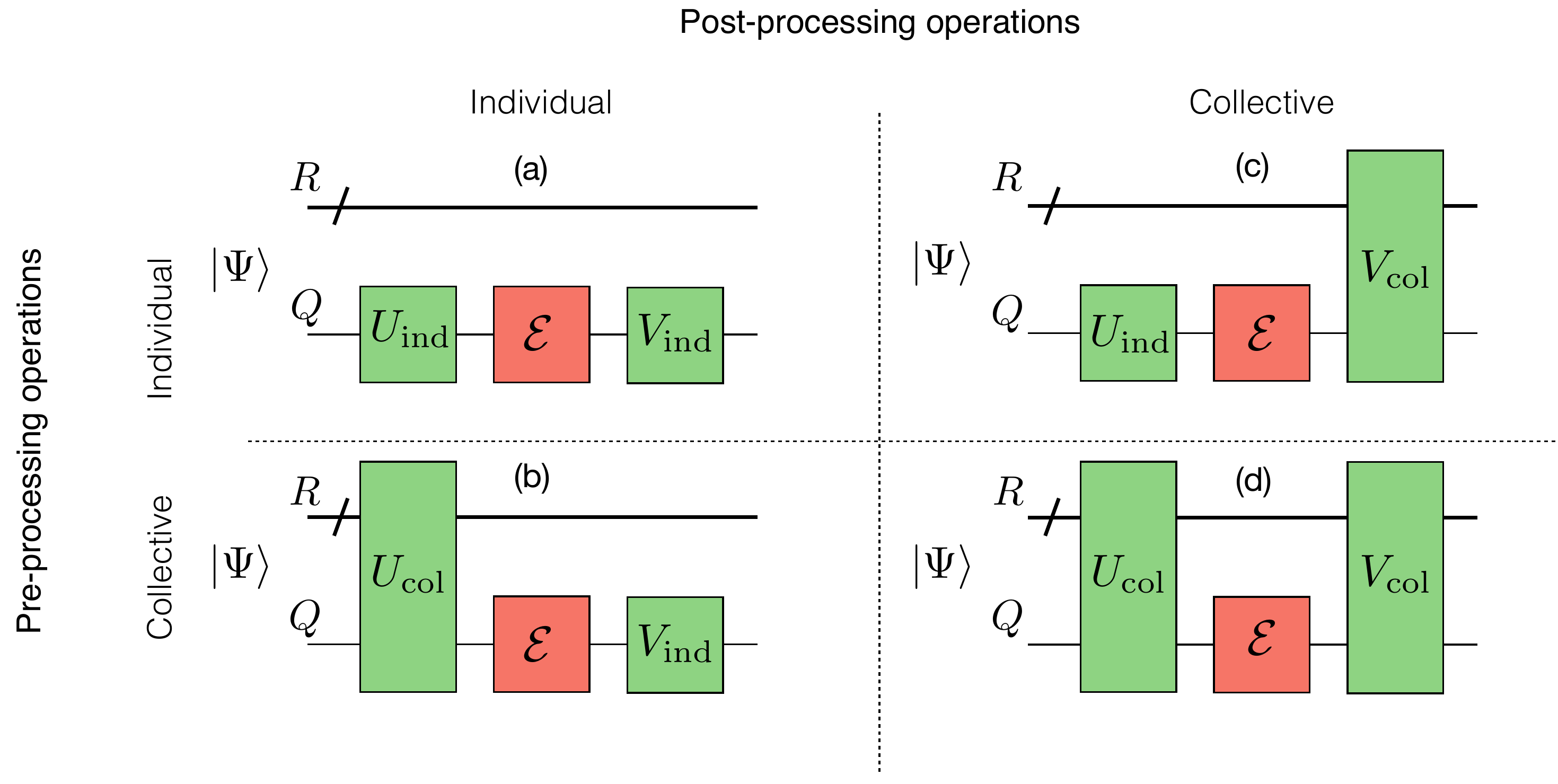}
		\vskip -4mm
		\caption{
			Schemes for protecting a pure state $\ket{\Psi}_{QR}$ from the decoherence channel $\mathcal{E}$, which acts on a particular qubit $Q$ by employing unitary operations.
			In (a) `both individual' scheme with both unitary operators are individual and they act on $Q$ is presented. 
			In (b) `collective-then-individual' scheme with the first operator is individual, while the second is non-individual, is shown.
			In (c) `individual-then-collective' scheme with the first operator is collective, while the second is individual, is presented. 
			In (d) the `both collective' scheme with the both operators are non-individual is shown.}
		\label{fig:schemes}
	\end{figure*}

	The initial reduced state of qubit $Q$ is as follows:
	\begin{equation}
	\rho_Q := {\rm tr}_R\ket{\Psi}_{QR}\bra{\Psi} = \sum_{i=0,1}\lambda_i\ket{\psi_i}_Q\bra{\psi_i}.
	\end{equation}
	The considered entropies for the state $\rho_Q$ then take the following forms:
	\begin{eqnarray}
	& S_{\rm vN}^Q:= S_{\rm vN}(\rho_Q) =-\lambda_0\log\lambda_0-(1-\lambda_0)\log(1-\lambda_0), \label{eq:svN} \\
	& S_{\rm lin}^Q:= S_{\rm lin}(\rho_Q) = 1-\sum_{i=0,1}\lambda_i^2=2\lambda_0(1-\lambda_0), \label{eq:slin} \\
	& S_{\rm min}^Q := S_{\min}(\rho_Q) = -\log\lambda_0. \label{eq:smin}
	\end{eqnarray}
	All these quantities can be expressed via single parameter $\lambda_0$ only, and they monotonously decrease with $\lambda_0$ as it ranges from $1/2$ to $1$.
	Thus, there are one-to-one correspondences between linear entropy, von Neumann entropy, and min-entropy for the qubit case.
	We also note that for the pure state $\ket{\Psi}_{QR}$ the von Neumann entropy $S_{\rm vN}^Q$ equals to the entropy of entanglement and the quantum discord~\cite{Ollivier2001,Henderson2001,Kiktenko2014,Kiktenko2015} 
	with respect to the $R-Q$ partitioning.
	
	Below all four strategies for our protection scheme that depicted in Fig.~\ref{fig:schemes} are considered, and  corresponding values of maximal achievable fidelity are derived.
	
	\subsection{`Both individual' scheme}
	
	Here we consider the scheme depicted in Fig.~\ref{fig:schemes}(a) step by step.
	First, we apply, as a pre-processing procedure, the following individual unitary operation $U_{\rm ind}$:
	\begin{equation}~\label{eq:U}
	U_{\rm ind} = \sum_{i=0,1}\ket{\phi_i}\bra{\psi_i},
	\end{equation}
	where $\{\ket{\phi_i}\}$ is a new orthonormal basis in the qubit space.
	After acting operation $U_{\rm ind}$, the state of the whole system is as follows:
	\begin{equation}\label{eq:stateboth}
	\ket{\Phi}_{QR} = \sum_{i=0,1} \sqrt{\lambda_i} \ket{\phi_i}_Q\otimes\ket{\zeta_i}_R.
	\end{equation}
	
	Then, after the action of the decoherence channel $\mathcal{E}$, state \eqref{eq:stateboth} transforms as follows: 
	\begin{equation}\label{eq:rhodec}
	\rho^{\rm dec}_{QR} = \sum_{i,j} \mathcal{E}[\ket{\phi_i}_Q\bra{\phi_j}]\otimes\ket{\zeta_i}_R\bra{\zeta_j}.
	\end{equation}
	
	As a post-processing procedure, we apply the second individual unitary operator $V_{\rm ind}$ of the following form:
	\begin{equation}
	V_{\rm ind}  = \sum_{i=0,1}\ket{\psi_i}\bra{\chi_i},
	\end{equation}
	where $\{\ket{\chi_i}\}$ is another orthonormal basis. 
	
	Using the fact that $\ket{\zeta_0}$ and $\ket{\zeta_1}$ are orthogonal and normalized, we obtain the following expression for the resulting fidelity:
	\begin{eqnarray}
	F_{\rm ind\;ind}=\bra{\Psi}_{QR}(V_{\rm ind}\otimes {\bf I}) \rho^{\rm dec}_{QR} (V_{\rm ind}^\dagger\otimes {\bf I})\ket{\Psi}_{QR}\\
	=\sum_{i,j}\lambda_i\lambda_j\bra{\chi_i}_Q\mathcal{E}[\ket{\phi_i}_Q\bra{\phi_j}]\ket{\chi_j}_Q,
	\end{eqnarray}
	where ${\bf I}$ stands for $2^{n-1}$ dimensional identity operator acting in the space of the subsystem $R$.
	
	The maximal value of the fidelity can be obtained by optimization over both unitary operators and can be written in the following form:
	\begin{equation}
	\begin{split}
	&\!\!\!\!\!F_{\rm ind\;ind}^{\rm opt} := \max_{U_{\rm ind}, V_{\rm ind}}  F_{\rm ind\;ind} \\ 
	&\!\!\!\!\!=\max_{\{ \ket{\phi_i} \}, \{ \ket{\chi_i} \}} \left(\sum_{i,j}\lambda_i\lambda_j\bra{\chi_i}_Q\mathcal{E}[\ket{\phi_i}_Q\bra{\phi_j}]\ket{\chi_j}_Q\right). 
	\end{split}
	\end{equation}
	We note that $F_{\rm ind\;ind}^{\rm opt}$ depends on $\{\lambda_i\}$ defining the initial reduced state of the qubit $Q$.
	
	\subsection{`Individual-then-collective' scheme}
	
	The second scheme, which is depicted in Fig.~\ref{fig:schemes}(b), begins with applying the individual unitary operator $U_{\rm ind}$ and decoherence channel $\mathcal{E}$ in the same way as in the `both individual' scheme.
	We then obtain the same state of the system given by Eq.~\eqref{eq:rhodec}.
	We can write its spectral decomposition given by the following expression:
	\begin{equation}\label{eq:spectrdec}
	\rho^{\rm dec}_{QR} = \sum_{i=0}^{2^n-1} p_i \ket{\xi_i}_{QR}\bra{\xi_i},
	\end{equation}
	where  $\{\ket{\xi_i}\}_{i=0}^{2^n-1}$ forms an orthonormal basis in whole $2^n$-dimensional Hilbert space, and we assume that $p_i\leq p_j$ for $i>j$.
	Since we are able to apply an arbitrary unitary transformation $V_{\rm col}$ to the whole state of the system, we can achieve the maximal possible fidelity (for a fixed form of $U_{\rm ind}$) equal to the largest eigenvalue of  $\rho^{\rm dec}_{QR}$.
	It can be written using min-entropy of $\rho^{\rm dec}_{QR}$ in the following form:
	\begin{equation}
	F_{\rm ind\;col}^{\rm prelim} = p_0 = \Lambda_{\max}(\rho^{\rm dec}_{QR})= 2^{-S_{\rm min}(\rho^{\rm dec}_{QR})}.
	\end{equation}
	This value of the fidelity can be achieved with $V_{\rm col}$ transforming the pure state with the highest eigenvalue $\ket{\xi_0}_{QR}$ into the initial state $\ket{\Psi}_{QR}$.
	Thus, we arrive to the optimal form of $V_{\rm col}$ given by:
	\begin{equation}
	V_{\rm col}^{\rm opt} = \ket{\Psi}\bra{\xi_0 } + [\ldots],
	\end{equation}
	where $[\ldots]$ stands for any appropriate remaining part of the unitary operator.
	
	The maximal value of the fidelity for the whole `individual-then-collective' scheme can be obtained by remaining optimization over $U_{\rm ind}$:
	\begin{equation}\label{eq:findcol}
	\begin{split}
	F_{\rm ind\;col}^{\rm opt}&:=\max_{U_{\rm ind}}F_{\rm ind\;col}^{\rm prelim} =\max_{\{\ket{\phi_i}\}}\left[2^{- S_{\rm min}(\rho_{QR}^{\rm dec})}\right] \\
	&=\exp\left[-\ln(2) \min_{\{\ket{\phi_i}\}}S_{\rm min}(\rho_{QR}^{\rm dec})\right].
	\end{split}
	\end{equation}
	
	\subsection{`Collective-then-individual' scheme}
	
	The third scheme is shown in Fig.~\ref{fig:schemes}(c) can be considered as the second scheme in the reverse order.
	To study this scheme it is useful to write the action of the map $\mathcal{E}$ via Kraus operators as follows:
	\begin{equation} \label{eq:Kraus}
	\mathcal{E}[\rho]=\sum_k A_k\rho A_k^\dagger,
	\end{equation} 
	where $\{A_k\}$ obeys the standard CPTP condition $\sum_k A_k^\dagger A_k={\bf 1}$, where ${\bf 1}$ stands for identity operator in the qubit space.
	
\begin{figure}
		\includegraphics[width=0.7\linewidth]{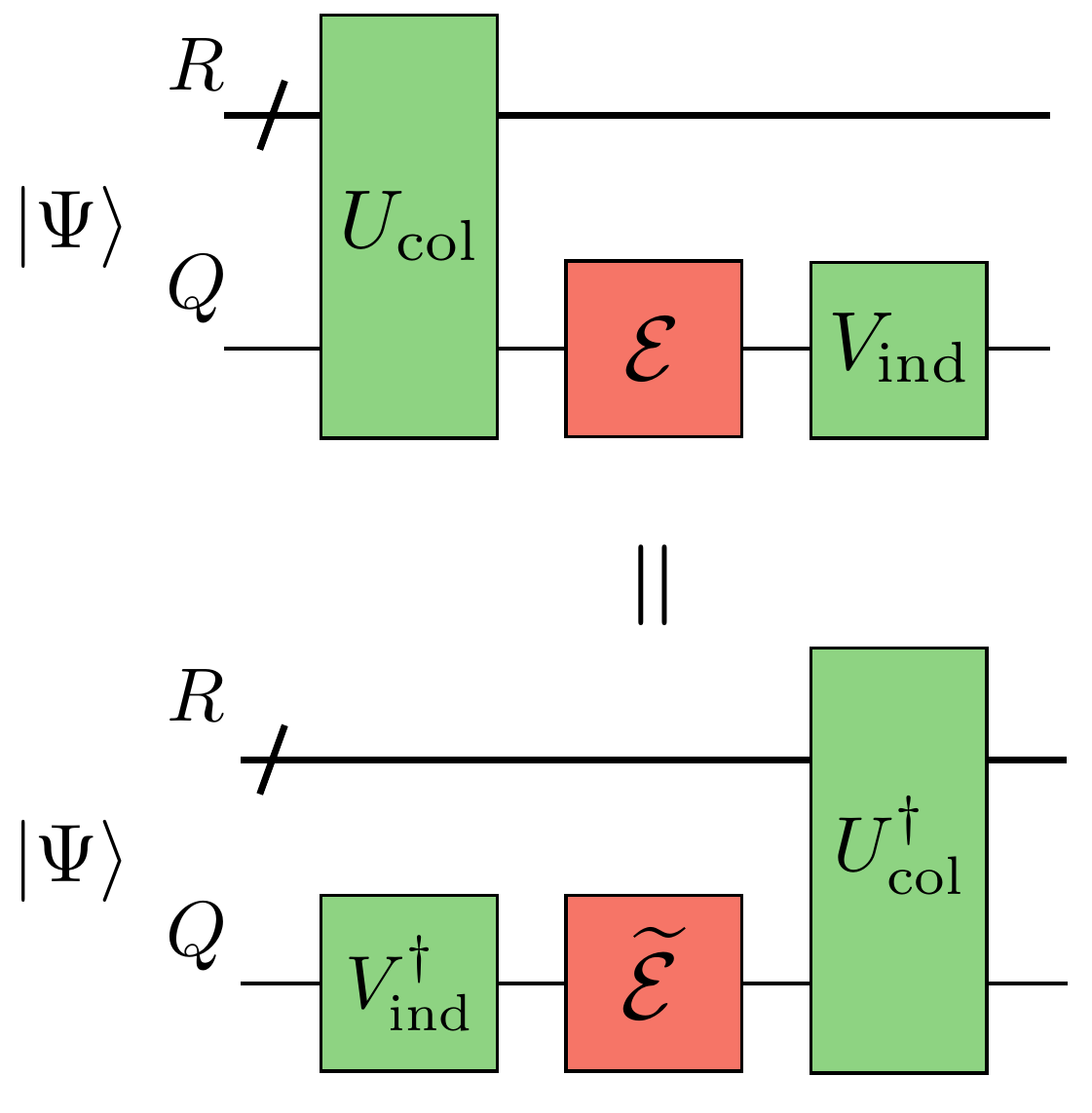}
		\vskip -4mm
		\caption{Equivalence between `collective-then-individual' and `individual-then-collective' schemes.}
		\label{fig:equiv}
	\end{figure}

	The resulting fidelity for some $U_{\rm col}$ and $V_{\rm ind}$ takes the following form:
	\begin{multline}
	F_{\rm col\;ind}=\sum_k\bra{\Psi}_{QR}(V_{\rm ind}\otimes {\bf I})A_kU_{\rm col}\ket{\Psi}_{QR}\bra{\Psi} \times \\U_{\rm col}^\dagger A_k^\dagger (V_{\rm ind}^\dagger\otimes {\bf I})\ket{\Psi}_{QR}=\sum_k
	\bra{\Psi}_{QR} U_{\rm col}^\dagger A_k^\dagger (V_{\rm ind}^\dagger\otimes {\bf I})\times\\\ket{\Psi}_{QR}\bra{\Psi}
	(V_{\rm ind}\otimes {\bf I})A_kU_{\rm col}\ket{\Psi}_{QR}	.
	\end{multline}
	One can see that the resulting expression is the same as the fidelity in the `individual-then-collective' scheme up to the following change (see Fig.~\ref{fig:equiv}):
	\begin{equation}\label{eq:symmetry}
		V_{\rm col} \leftrightarrow U_{\rm col}^\dagger, \quad U_{\rm ind} \leftrightarrow V_{\rm ind}^\dagger, \quad \mathcal{E}\leftrightarrow \widetilde{\mathcal{E}},
	\end{equation}
	where $\widetilde{\mathcal{E}}$ is a conjugate map to $\mathcal{E}$ of the following form:
	\begin{equation}\label{eq:Kraus_dual}
	\widetilde{\mathcal{E}}[\rho]:=\sum_k A_k^\dagger\rho A_k.
	\end{equation}	
	The maximal fidelity in the `collective-then-individual' scheme is given by the similar expression as in Eq.~\eqref{eq:findcol}:
	\begin{equation} \label{eq:fcolind}
	F_{\rm col\;ind}^{\rm opt}:=\exp\left[-\ln(2) \min_{\{\ket{\phi_i}\}}S_{\rm min}(\widetilde{\rho}_{QR}^{\rm dec})\right],
	\end{equation}
	where
	\begin{equation}
	\widetilde{\rho}_{QR}^{\rm dec} := \sum_{i,j} \widetilde{\mathcal{E}}[\ket{\phi_i}_Q\bra{\phi_j}]\otimes\ket{\zeta_i}_R\bra{\zeta_j}.
	\end{equation}
	
	We then show that the maximal fidelities is given by Eq.~\eqref{eq:fcolind} and Eq.~\eqref{eq:findcol} are the same.
	Let $\ket{\xi^{\rm opt}}$ and $\{\ket{\phi_i^{\rm opt}}\}$ are such that $F^{\rm opt}_{\rm ind\;col}$ achieves its maximal value of the following form:
	\begin{equation}\label{eq:findcolnew}
	\begin{split}
	&\!\!\!\!\!\!\!\!\!F^{\rm opt}_{\rm ind\;col}= \sum_{k,i,j} \lambda_i\lambda_j \times \\
	&\!\!\!\!\!\!\!\!\!{\times}\bra{\xi^{\rm opt}}_{QR} \left( A_k \ket{\phi_i^{\rm opt}}\bra{\phi_j^{\rm opt}} A_k^\dagger \otimes \ket{\zeta_i}_R\bra{\zeta_j} \right) \ket{\xi^{\rm opt}}_{QR}.
	\end{split}
	\end{equation}
	One can see that $\ket{\xi^{\rm opt}}_{QR}$ is an eigenvector corresponding to the largest eigenvalue of $\rho_{QR}^{\rm dec}$ obtained after optimal individual operation defined by $\{\ket{\phi_i^{\rm opt}}\}$.
	
	We introduce matrix elements of Kraus operators in the basis of $\{\ket{\phi_i^{\rm opt}}\}$:
	\begin{equation}
	a^k_{ij} := \bra{\phi_i^{\rm opt}} A_k \ket{\phi_j^{\rm opt}},
	\end{equation}
	and also introduce the overlap between $\ket{\xi^{\rm opt}}$ and the vector set $\{\ket{\phi_i^{\rm opt}}\otimes\ket{\zeta_j}\}_{ij}$:
	\begin{equation}
	C_{ij} := \left(\bra{\phi_i^{\rm opt}} \otimes \bra{\zeta_j}\right)\ket{\xi^{\rm opt}}.
	\end{equation}
	Then expression \eqref{eq:findcolnew} takes the following form:
	\begin{equation}
	\begin{split}
	F^{\rm opt}_{\rm ind\;col}&=\sum_{k,i,j,m,n}\lambda_i\lambda_j C_{mi}^*a^k_{mi}a^{k*}_{nj} C_{nj} \\
	&=\sum_k\left(\sum_{i,m}\lambda_iC_{mi}^*a^k_{mi}\right)\left(\sum_{j,n}\lambda_jC_{nj}a^{k*}_{nj} \right) \\
	&=\sum_k \left|\sum_{i,m} \lambda_i C_{mi}^*a^k_{mi} \right|^2.
	\end{split}
	\end{equation}
	
	\begin{figure*}
		\includegraphics[width=0.7\linewidth]{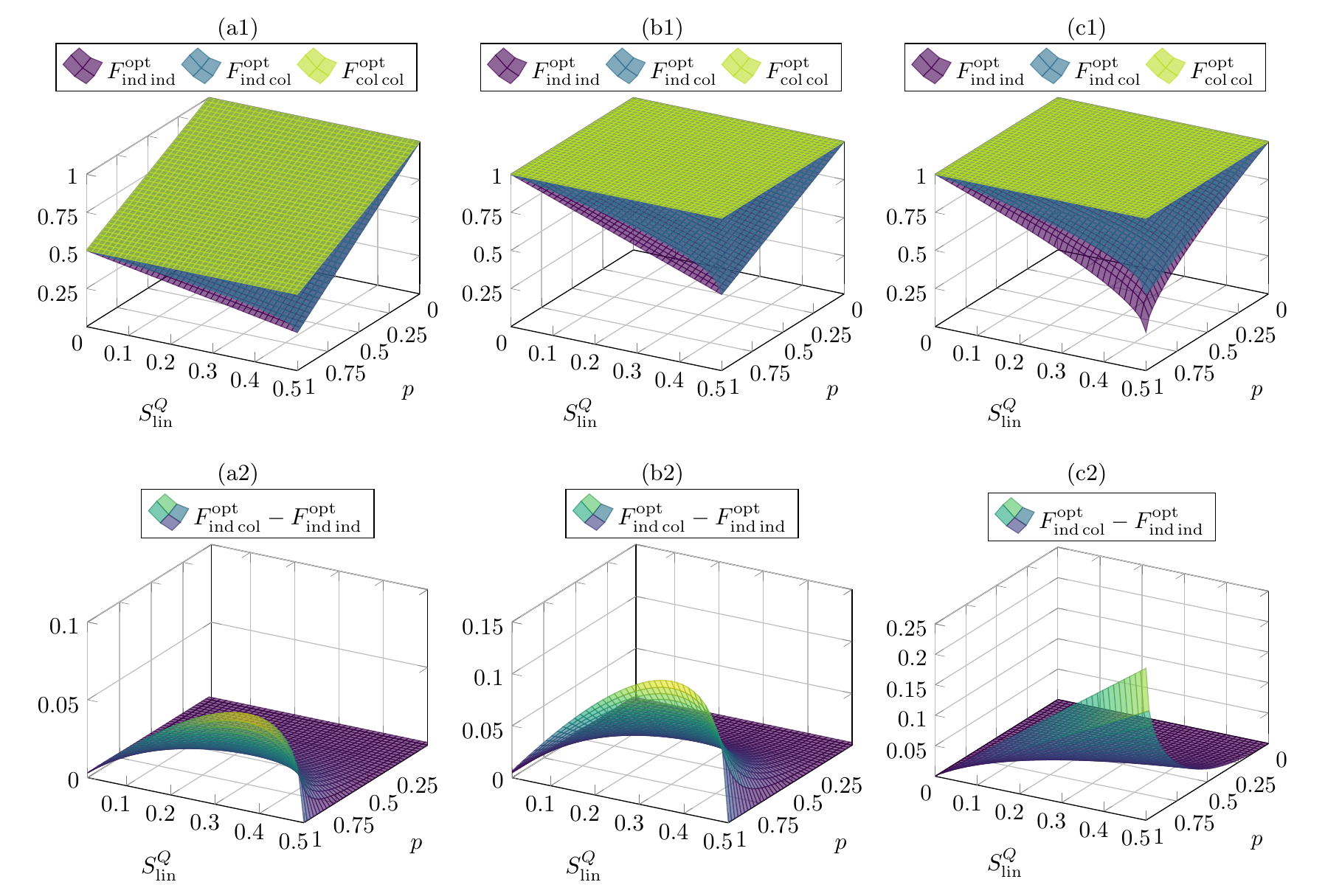}
		\vskip -4mm
		\caption{The results of employing the proposed protecting schemes.
		In (a) results for depolarizing channels, in (b) results for dephasing channels, are in (c) results for amplitude damping channels are presented.
		In (a1), (b1), and (c1) the maximal fidelities for all different strategies are shown.
		In (a2), (b2), and (c2) the advantage of using `individual-then-collective' over `both individual' is demonstrated.
		We note that `individual-then-collective' and `collective-then-individual' provides the same results.}
		\label{fig:3d}
	\end{figure*}
	
	Let us then return to the `collective-then-individual' scheme.
	As we demonstrated above, it is equivalent to `individual-then-collective' with the only difference that the map $\mathcal{E}$ is changed by the dual map~$\mathcal{\widetilde{E}}$.
	One can write the fidelity of the `collective-then-individual' scheme in the form similar to Eq.~\eqref{eq:findcolnew}:
	\begin{equation} \label{eq:fcolindnew}
	\begin{split}
		&\!\!\!\widehat{F}_{\rm col\;ind}= \\ 
		&\!\!\!\sum_{k,i,j} \lambda_i\lambda_j
	\bra{\widetilde{\xi}}_{QR} \left( A_k^\dagger \ket{\widetilde{\phi}_i}\bra{\widetilde{\phi}_j} A_k \otimes \ket{\zeta_i}_R\bra{\zeta_j} \right) \ket{\widetilde{\xi}}_{QR},
	\end{split}
	\end{equation}
	where $\ket{\widetilde{\xi}}$ is some $2^n$-dimensional normalized vector and $\{ \ket{\widetilde{\phi}_i} \}$ is an orthonormal basis in qubit space.
	We can set
	\begin{equation}
	\ket{\widetilde{\phi}_i} := \ket{{\phi}_i^{\rm opt}},
	\end{equation}
	and choose $\ket{\widetilde{\xi}}$ such that
	\begin{equation}
	\left(\bra{\phi_i^{\rm opt}} \otimes \bra{\zeta_j}\right)\ket{\widetilde{\xi}} = C_{ji}^*.
	\end{equation}
	Then we obtain the following expression for the fidelity:
	\begin{equation}
	\begin{split}
	\widehat{F}_{\rm col\;ind}=\sum_{k,i,j,m,n} \lambda^{}_i\lambda^{}_j C_{im} a_{im}^{*k} a_{jn}^k C_{jn}^* \\
	=\sum_k\left|\sum_{j,n} \lambda^{}_j a_{jn}^k C_{jn}^* \right| = F^{\rm opt}_{\rm ind\;col}.
	\end{split}
	\end{equation}
	On the one hand, we have the following expression:
	\begin{equation}
		F^{\rm opt}_{\rm col\;ind}\geq\widehat{F}_{\rm col\;ind}= F^{\rm opt}_{\rm ind\;col}.
	\end{equation}
	On the other hand, by employing a similar line reasoning we can obtain $F^{\rm opt}_{\rm ind\;col}\geq F^{\rm opt}_{\rm col\;ind}$.
	Finally, we arrive the following relation: $F^{\rm opt}_{\rm col\;ind}=F^{\rm opt}_{\rm ind\;col}$, i.e. there is a full equivalence of the performance of both schemes.
	The observed identity reveals a time-symmetry aspect of transformation~(\ref{eq:symmetry}).
	That is why are all our subsequent considerations we will take into account the `individual-then-collective' scheme only since all the results for `collective-then-individual' scheme can be obtained from it straightforwardly.
	
	\subsection{`Both collective' scheme}
	
	Finally, we consider a scheme, which is presented in Fig.~\ref{fig:schemes}(d).
	Although this scheme may seem far from practical application, it is important from a theoretical point of view since it provides an upper bound on the fidelity level achieved in the considered class of schemes.
	
	We can use any operator $U_{\rm col}$ in order to `re-prepare' any desired state before acting of the decoherence channel. 
	In this way, the optimal strategy is to choose $U_{\rm col}$ such that the resulting pure state is affected by $\mathcal{E}$ as less as possible.
	Since $\mathcal{E}$ acts individually on $Q$, the best fidelity can be achieved with $U_{\rm col}$ transforming $\ket{\Psi}_{QR}$ in a separable state of the following form:
	\begin{equation}
	U_{\rm col}\ket{\Psi}_{QR} = \ket{\Upsilon}_Q\otimes\ket{\Xi}_R,
	\end{equation}
	with some normalized pure states $\ket{\Upsilon}_Q$ and $\ket{\Xi}_R$ in 2-dimensional and $2^{n-1}$-dimensional Hilbert spaces correspondingly. 
	
	Since the second unitary $V_{\rm col}$ can turn any pure state into $\ket{\Psi}$, the best final fidelity will be achieved with $\ket{\Upsilon}$ minimizing a min-entropy of $\mathcal{E}[\ket{\Upsilon}\bra{\Upsilon}]$.
	Denote
	\begin{equation}
	\ket{\Upsilon^{\rm opt}} := \arg\min_{\ket{\Upsilon}}S_{\rm min}(\mathcal{E}[\ket{\Upsilon}\bra{\Upsilon}]).
	\end{equation}
	We note that the minimum for min-entropy is the same a minimum for linear and von Neumann entropies since $\mathcal{E}$ is a qubit channel:
	\begin{equation}
	\ket{\Upsilon^{\rm opt}} = \arg\min_{\ket{\Upsilon}}S_{\rm lin}(\mathcal{E}[\ket{\Upsilon}\bra{\Upsilon}])\\=\arg\min_{\ket{\Upsilon}}S_{\rm vN}(\mathcal{E}[\ket{\Upsilon}\bra{\Upsilon}]).
	\end{equation}
	We then obtain the following expression for the maximal fidelity as follows:
	\begin{equation}
	\begin{split}
	F_{\rm col\;col}^{\rm opt}&= \Lambda_{\max}(\mathcal{E}[\ket{\Upsilon^{\rm opt}}\bra{\Upsilon^{\rm opt}}])\\
	&=\exp\left[\ln(2)S_{\min}(\mathcal{E}[\ket{\Upsilon^{\rm opt}}\bra{\Upsilon^{\rm opt}}]) \right].
	\end{split}
	\end{equation}
	This value is achieved under the action of unitary operators of the following form:
	\begin{eqnarray}
	&&U_{\rm col} = \ket{\Upsilon^{\rm opt}}\otimes\ket{\Xi}\bra{\Psi}+[\ldots],\\
	&&V_{\rm col} = \ket{\Psi}\bra{\Theta}\otimes\bra{\Xi}+[\ldots],
	\end{eqnarray}
	where $\ket{\Theta}$ stands for an eigenstate of $\mathcal{E}[\ket{\Upsilon^{\rm opt}}\bra{\Upsilon^{\rm opt}}]$ corresponding to largest eigenvalue $\Lambda_{\max}(\mathcal{E}[\ket{\Upsilon^{\rm opt}}\bra{\Upsilon^{\rm opt}}])$ 
	and $[\ldots]$ stands for the arbitrary appropriate remaining part of the unitary operators.
	
	In conclusion of the section, we note the following relation between all the considered approaches:
	\begin{equation}
	F_{\rm ind\;ind}^{\rm opt}\leq F_{\rm col\;ind}^{\rm opt} =  F_{\rm ind\;col}^{\rm opt} \leq F_{\rm col\;col}^{\rm opt}.
	\end{equation}
	It holds true since the scheme with two individual operations is a particular case for individual-then-collective (or collective-then-individual) scheme, while the latter schemes are particular cases for the scheme with two collective operations.
	
	\section{Applications for the basic decoherence channels}\label{sec:models}
		
	Here we apply the described above schemes to main decoherence models, particularly depolarizing, dephasing, and amplitude damping quantum channels~\cite{Nielsen}.
	We derive the values of maximal achievable fidelities depending on the strength of particular decoherence channel and the structure of the reduced state $\rho_Q$ of the qubit affected by this channel.
	Without loss of generality, we consider the initial state  $\ket{\Psi}_{QR}$ of the quantum system in the following form:
	\begin{equation}\label{eq:Psiin}
	\ket{\Psi}_{QR} = \sum_{i=0,1} \sqrt{\lambda_i} \ket{i}_Q\otimes\ket{\zeta_i}_R,
	\end{equation}
	where $\ket{0}$ and $\ket{1}$ form a standard computational basis.
	It can be done since one can always add an individual unitary operator to the pre-processing operator, which rotates the eigenstates of the reduced state $\rho_Q$ to any set of orthonormal states.
	
	\subsection{Depolarizing channel}
	
	The action of depolarizing channel on an arbitrary qubit state $\rho$ is as follows:
	\begin{equation}
	\mathcal{E}[\rho] = (1-p)\rho + p\frac{{\bf 1}}{2}{\rm tr}(\rho),
	\end{equation}
	where ${\bf 1}$ is the two-dimensional identity matrix and $p\in[0,1]$ is a parameter describing a depolarization strength.
	It can be interpreted as turning the state into the maximally mixed state ${\bf 1}/2$ with probability $p$, and leaving it untouched with probability $1-p$.
	
	\subsubsection{`Both individual' scheme}
	For any qubit unitary operator $u$, we have the following property of the depolarizing channel:
	\begin{equation} \label{eq:depolprop}
	\mathcal{E}[u\rho u^\dagger]=u\mathcal{E}[\rho]u^\dagger.
	\end{equation}
	That is why one can set $U_{\rm ind}:={\bf 1}$ in `both individual' as well as `individual-then-collective' schemes, since all the necessary operations can be releazed with the use of post-processing unitary operator.
	
	The state after decoherence takes the following form:
	\begin{equation}\label{eq:rhodecdepol}
	\rho^{\rm dec}_{QR}=(1-p)\ket{\Psi}_{QR}\bra{\Psi} + p \frac{\bf 1}{2}\otimes \sum_i\lambda_i\ket{\zeta_i}_R\bra{\zeta_i}.
	\end{equation}
	One can easily check that the maximal fidelity in the `both individual' scheme is achieved with $V_{\rm ind}={\bf 1}$, and it is as follows:
	\begin{equation} \label{eq:Findinddepol}
	F_{\rm ind\;ind}^{\rm opt}= (1-p) + \frac{p}{2}\sum_i\lambda_i^2=1-\frac{p}{2}(1+S_{\rm lin}^Q),
	\end{equation}
	where $S_{\rm lin}^Q$ is a linear entropy of the reduced state of $Q$ in the initial state, see Eq.~\eqref{eq:slin}.
	Thus, we obtain that the fidelity decreases linearly with the growth of decoherence strength and linear entropy as it is shown in Fig.~\ref{fig:3d}(a1).
	
	\subsubsection{`Individual-then-collective' scheme}
	
	The state after the depolarizing channel in the `individual-then-collective' scheme with  $U_{\rm ind}:={\bf 1}$ is given by Eq.~\eqref{eq:rhodecdepol}.
	The optimal fidelity in the considered scheme equals to its largest eigenvalue and it is as follows:
	\begin{equation}
	F_{\rm ind\;col}^{\rm opt} = \frac{1}{2}-\frac{1}{4} p+\frac{1}{4} \sqrt{(p-2)^{2}-2 p(4-3 p) S^Q_{\rm lin}}.
	\end{equation}
	The behaviour the fidelity $F_{\rm ind\;col}^{\rm opt}$ is presented in Fig.~\ref{fig:3d}(a1), while the difference $F_{\rm ind\;col}^{\rm opt}-F_{\rm ind\;ind}^{\rm opt}$ is given in Fig.~\ref{fig:3d}(a2).
	We note that both fidelities $F_{\rm ind\;ind}^{\rm opt}$ and $F_{\rm ind\;col}^{\rm opt}$ decrease with increasing of $S_{\rm lin}^Q$.
	The efficiency of employing a single collective operator over two individual operators growth with an increase of $p$ and it is maximal for $p=1$ and $S_{\rm lin}^Q \approx 0.375$.
	
\begin{figure}
		\includegraphics[width=0.9\linewidth]{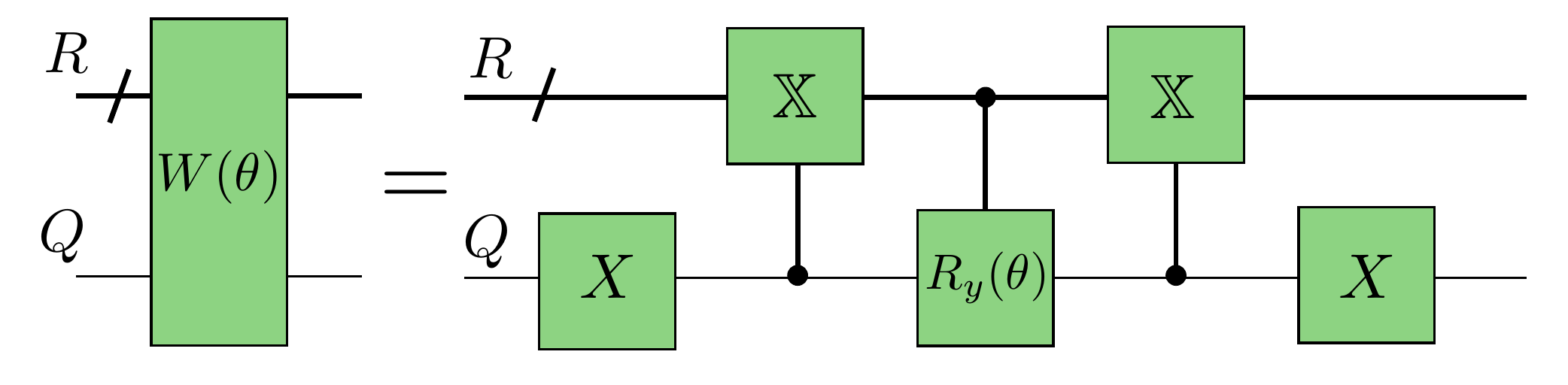}
		\caption{Decomposition of the collective unitary operation $W$ given by Eq.~\eqref{eq:vnocol} into elementary operations.}
		\label{fig:nocol}
	\end{figure}

	The subtle question is how to obtain the optimal form of the post-processing operator $V_{\rm col}$.
	This operator has to turn the eigenstate $\ket{\xi_0}_{QR}$, which corresponds to the largest eigenvalue of $\rho^{\rm dec}_{QR}$ to the initial state $\ket{\Psi}_{QR}$.
	For the state given by Eq.~\eqref{eq:rhodecdepol}, we have the following representation of $\ket{\xi_0}_{QR}$ in the basis 
	$\{\ket{0}_Q\otimes\ket{\zeta_0}_R, \ket{0}_Q\otimes\ket{\zeta_1}_R, \ket{1}_Q\otimes\ket{\zeta_0}_R, \ket{1}_Q\otimes\ket{\zeta_1}_R\}$:
	\begin{equation}\label{eq:xiform}
	\ket{\xi_0}_{QR} =
	\begin{bmatrix}
	\cos\upsilon \\ 0 \\ 0 \\ \sin\upsilon
	\end{bmatrix}
	\end{equation}
	with
	\begin{equation}
	\upsilon = \arctan\frac{2F_{\rm ind\;col}-(2-p)\lambda_0}{2(1-p)\sqrt{S^Q_{\rm lin}/2}}.
	\end{equation}
	In order to transform $\ket{\xi}_{QR}$ into $\ket{\Psi}_{QR}$, which in the same basis has the following form:
	\begin{equation}
	\ket{\Psi}_{QR}=\begin{bmatrix}
	\sqrt{\lambda_0} \\ 0 \\ 0 \\ \sqrt{\lambda_1},
	\end{bmatrix},
	\end{equation}
	one can employ the operator as follows:
	\begin{equation} \label{eq:Vcoldepol}
	V_{\rm col}=W(2\gamma),
	\end{equation}
	where
	\begin{equation} \label{eq:vnocol}
	W(\theta):=
	\begin{bmatrix}
	\cos(\theta/2) &  0 & 0 & \sin(\theta/2) \\
	0 & 1 & 0 & 0 \\
	0 & 0 & 1 & 0 \\
	-\sin(\theta/2) & 0 & 0 & \cos(\theta/2)
	\end{bmatrix}.
	\end{equation}
	with
	\begin{equation} \label{eq:varpi}
	\gamma := -\arccos \bra{\xi_0} \Psi\rangle\\=-\arccos( \sqrt{\lambda_0}\cos\upsilon+\sqrt{\lambda_1}\sin\upsilon).
	\end{equation}
	The operator $W(\theta)$ can be constructed with a circuit presented in Fig.~\ref{fig:nocol}, where we use the following notations:
	\begin{equation}
	X := \ket{1}\bra{0}+\ket{0}\bra{1}, \quad \mathbb{X} := \ket{\xi_1}\bra{\xi_0}+\ket{\xi_0}\bra{\xi_1}.
	\end{equation}
	The rotation operation around $y$-axis of the Bloch sphere,
	\begin{equation}
	R_y(\theta)=\begin{bmatrix}
	\cos(\theta/2) &  -\sin(\theta/2) \\
	\sin(\theta/2) & \cos(\theta/2)
	\end{bmatrix},
	\end{equation}
	in the central controlled-unitary gate is assumed to be performed if the subsystem $R$ is in the state $\ket{\zeta_1}$.
	
	\subsubsection{`Both collective' scheme}
	Since in the case of depolarizing channel the action of individual unitary operation cannot affect the resulting entropy, as it follows from the property Eq.~\eqref{eq:depolprop}, we can set the pre-processing unitary as follows:
	\begin{equation} \label{eq:ucoldep}
	U_{\rm col} := W(2\varsigma), \quad \varsigma := \arctan\sqrt{\lambda_1/\lambda_0},
	\end{equation}
	which turn the state $\ket{\Psi}_{QR}$ into the state $\ket{0}_Q\ket{\xi_0}_R$.
	Then after acting the depolarizing channel, we obtain:
	\begin{equation}
	\!\!\!\rho_{QR}^{\rm dec} = \left((1-\frac{p}{2}) \ket{0}_Q\bra{0}_Q+\frac{p}{2} \ket{1}_Q\bra{1}_Q\right)\otimes \ket{\xi_0}_R\bra{\xi_0},
	\end{equation}
	with the largest eigenvalue equal to $1-p/2$ and corresponding eigenvector $\ket{0}_Q\ket{\xi_0}_R$.
	Thus, finally, we have the following expression:
	\begin{equation}
	F_{\rm col\;col}^{\rm opt}=1-\frac{p}{2},
	\end{equation}
	It is achieved if the operators are as follows:
	\begin{equation} \label{eq:vcoldep}
	V_{\rm col} := U_{\rm col}^\dagger =  W(-2\varsigma),
	\end{equation}
	where $\varsigma$ is given by Eq.~\eqref{eq:ucoldep}.
	
	The unavoidable decrease of the fidelity can be explained by an unavoidable entropy production effect inherent to the depolarizing channel.
	The behaviour of $F_{\rm col\;col}^{\rm opt}$ is also presented in Fig.~\ref{fig:3d}(a1).
	
	\subsection{Dephasing channel}
	
	The action of the dephasing channel on a qubit state $\rho$ has the following form:
	\begin{equation}
	\mathcal{E}[\rho] = (1-p)\rho + p\left( \rho_{00}\ket{0}\bra{0}+\rho_{11} \ket{1}\bra{1} \right),
	\end{equation}
	where $\rho_{ii}:=\bra{i}\rho\ket{i}$ are diagonal elements of $\rho$ and $p\in[0,1]$ is again a strength of decoherence.
	One can note that the dephasing channel destroys non-diagonal elements of the density matrix $\rho$ with probability $p$ and does not change the state with probability $1-p$.
	In contrast to the depolarizing channel, we see that the dephasing channel does not affect the diagonal states.
	This feature can be efficiently employed in protection schemes.
	
	\subsubsection{`Both individual' scheme}
	
	It is easy to check that the optimal fidelity for the `both individual' scheme is achieved for	$U_{\rm ind}=V_{\rm ind}={\bf 1}$, since these operations left the reduced state of $Q$ in the `safe' diagonal form.
	For the identity pre-processing operator, we obtain the state after dephasing as follows:
	\begin{equation} \label{eq:dephstate}
	\rho_{QR}^{\rm dec}=(1-p)\ket{\Psi}_{QR}\bra{\Psi} + p\sum_{i}\lambda_i\ket{i}_Q\bra{i}\otimes\ket{\xi}_R\bra{\xi}.
	\end{equation}
	After applying the identity post-processing the value of fidelity takes the form: 
	\begin{equation}
	F_{\rm ind\;ind}^{\rm opt}=(1-p)+p\sum_i\lambda_i^2=1-pS_{\rm lin}^Q.
	\end{equation}
	One can see that in the case of the depolarizing channel \eqref{eq:Findinddepol} the fidelity in dephasing channel decreases linearly with a growth of $p$ and $S_{\rm lin}^Q$ (see also Fig.\ref{fig:3d}(b1)).
	
	\subsubsection{`Individual-then-collective' scheme}
	
	In the `individual-then-collective' scheme the optimal individual operator is again the identity operator $U_{\rm ind}:={\bf 1}$, which provides the state after decoherence given by Eq.~\eqref{eq:dephstate}.
	Its maximal eigenvalue determines the maximal fidelity as follows:
	\begin{equation}
	F_{\rm ind\;col}=\frac{1}{2}+\frac{1}{2} \sqrt{1-2 p(2-p) S_{\rm lin}^Q}
	\end{equation}
	and the corresponding eigenstate has the same form as in Eq.~\eqref{eq:xiform} with the only difference that
	\begin{equation} \label{eq:upsilondeph}
	\upsilon := \arctan\frac{F^{\rm opt}_{\rm ind\;col}-\lambda_0}{(1-p)\sqrt{S^Q_{\rm lin}/2}}.
	\end{equation}
	Thus, the optimal collective post-processing operator is given by Eq.~\eqref{eq:Vcoldepol} with $\gamma$ calculated using Eq.~\eqref{eq:varpi} with the updated value of $\upsilon$ from Eq.~\eqref{eq:upsilondeph}.
	
	We demonstrate the behaviour of the fidelity $F_{\rm ind\;col}$ in Fig.~\ref{fig:3d}(b1) together with the difference $F_{\rm ind\;col}-F_{\rm ind\;ind}$ in Fig.~\ref{fig:3d}(b2).
	One can see that the maximal difference in the case of the dephasing is in two times higher than the one for depolarizing channel and is achieved at the same point $p=1$, $S_{\rm lin}^Q \approx 0.375$.
	
	\subsubsection{`Both collective' scheme}
	
	One can see that the state $\ket{0}_Q\ket{\xi_0}_Q$ is not affected by dephasing channels, 
	that is why we can employ the pre-processing and post-processing collective operators in the forms given by Eq.~\eqref{eq:ucoldep} and Eq.~\eqref{eq:vcoldep}, correspondingly.
	They allow achieving the maximal fidelity 
	$ F_{\rm col\;col}^{\rm opt}=1$.
	
	\subsection{Amplitude damping channel}
	
	We complete our consideration of the basic decoherence models with an amplitude damping channel. 
	We can write its action in terms of Kraus operators:
	\begin{equation}
	A_1 = \begin{bmatrix} {1} &  {0} \\ {0} & {\sqrt{1-p}}\end{bmatrix}, \quad
	A_2 = \begin{bmatrix} {0} &  {\sqrt{p}} \\ {0} &  {0}\end{bmatrix}.
	\end{equation}
	Here $p\in[0,1]$ is again a decoherence strength.
	The amplitude damping process corresponds to the relaxation of the `excited' state $\ket{1}$ to the `ground' state $\ket{0}$.
	We note that there is an apparent asymmetry between the behaviour of $\ket{0}$ and $\ket{1}$: $\ket{0}$ remains untouched by the channel, while $\ket{1}$ undergoes a transformation to $\ket{0}$.
	Here we would like to remember that according to our convention we have $\lambda_0\geq\lambda_1$ in the initial state~\eqref{eq:Psiin}.
	It corresponds to the fact that we assume that $Q$ has the `ground' level to be populated more than the `excited' one.
	
	\begin{table*}[t]
		\begin{tabular}{p{0.16\linewidth}|p{0.27\linewidth}|p{0.37\linewidth}|p{0.12\linewidth}} \hline\hline
			& `Both individual' & `individual-then-collective' \& `collective-then-individual' & `Both collective' \\
			\hline
			Depolarizing & $1-p(1+S_{\rm lin}^Q)/2$ & $1/2- p/4+\sqrt{(p-2)^{2}-2 p(4-3 p) S^Q_{\rm lin}}/4$ & $1-p/2$ \\
			Dephasing & $1-pS_{\rm lin}^Q$ & $1/2+\sqrt{1-2 p(2-p) S_{\rm lin}^Q}/2$& 1 \\
			Amplitude damping & $\sqrt{1-p}S_{\rm lin}^Q+(1-p/2)(1-S_{\rm lin}^Q)+p\sqrt{1-2S_{\rm lin}^Q}/2$ & $1-p\left(1-\sqrt{1-2 S_{\rm lin}^Q}\right)/2$& 1 \\
			\hline\hline
		\end{tabular}
		\caption{The maximal achievable fidelities in all the considered approaches and decoherence models as functions of decoherence strength $p$ and initial linear entropy of the reduced state of the decohered qubit $S_{\rm lin}^Q$.}
		\label{tbl}
	\end{table*}
	
	\subsubsection{`Both individual' scheme}
	The optimal fidelity in the `both individual' scheme is achieved for the $U_{\rm ind}$ and $V_{\rm ind}$ being equal to identity operators: $U_{\rm ind}=V_{\rm ind}={\bf 1}$.
	The state after the decoherence takes the following form:
	\begin{equation} \label{eq:rhodecad}
	\rho_{QR}^{\rm dec}=\begin{bmatrix}
	{\lambda_{0}} & {0} & {0} & {\sqrt{\lambda_{0}} \sqrt{\lambda_{1}} \sqrt{\overline{p}}} \\ 
	{0} & {p \lambda_{1}} & {0} & {0} \\ 
	{0} & {0} & {0} & {0} \\ 
	{\sqrt{\lambda_{0}} \sqrt{\lambda_{1}} \sqrt{\overline{p}}} & {0} & {0} & {\overline{p} \lambda_{1}}
	\end{bmatrix},
	\end{equation}
	where $\overline{p}:=1-p$ and the matrix is written in the basis $\{\ket{0}_Q\otimes\ket{\zeta_0}_R, \ket{0}_Q\otimes\ket{\zeta_1}_R, \ket{1}_Q\otimes\ket{\zeta_0}_R, \ket{1}_Q\otimes\ket{\zeta_1}_R\}$.
	
	The resulting fidelity takes the following form:
	\begin{equation}
	F_{\rm ind\;ind}^{\rm opt} = \lambda_0^2+\lambda_1^2-\lambda_1^2p+2\lambda_0\sqrt{1-p}\lambda_1.
	\end{equation}
	By using the substitution,
	\begin{equation}
	\lambda_0 = \frac{1}{2}+\frac{1}{2}\sqrt{1-2S_{\rm lin}^Q},
	\end{equation}
	we obtain the following expression:
	\begin{equation}
	F_{\rm ind\;ind}^{\rm opt} =\sqrt{1-p}S_{\rm lin}^Q+\left(1-\frac{p}{2}\right)\left(1-S_{\rm lin}^Q\right)  
	+\frac{p}{2}\sqrt{1-2S_{\rm lin}^Q}
	\end{equation}
	We see that the fidelity decreases with the growth of $S_{\rm lin}^Q$ like in the case of depolarizing and dephasing channels, as it is shown in Fig.~\ref{fig:3d}(b1).
	
	\subsubsection{`Individual-then-collective' scheme}
	
	In the `individual-then-collective' scheme the best fidelity is also achieved for $U_{\rm ind}={\bf 1}$.
	The maximal fidelity is given by the largest eigenvalue of state~\eqref{eq:rhodecad}:
	\begin{equation}
	F_{\rm ind\;col}^{\rm opt}=1-p(1-\lambda_0) = 
	1-p\left(\frac{1}{2}-\frac{1}{2} \sqrt{1-2 S_{\rm lin}^Q}\right).
	\end{equation}
	The corresponding eigenstate has the same form as in Eq.~\eqref{eq:xiform} with
	\begin{equation} \label{eq:upsilondamp}
	\upsilon := \arctan\left(\lambda_1\sqrt{\frac{2(1-p)}{S_{\rm lin}^Q}}\right).
	\end{equation}
	The optimal collective post-processing operator is then given by the same expression~\eqref{eq:Vcoldepol} with $\gamma$ calculated by Eq.~\eqref{eq:varpi} with updated value of $\upsilon$ given by Eq.~\eqref{eq:upsilondamp}.
	
	We show the behaviour of $F_{\rm ind\;col}$ in Fig.~\ref{fig:3d}(c1).
	The corresponding difference $F_{\rm ind\;col}-F_{\rm ind\;ind}$ is presented in Fig.\ref{fig:3d}(c2).
	The maximal advantage of `individual-then-collective' ('collective-then-individual') scheme over `both individual' schemes in the case of the amplitude damping channel 
	is two times higher than the one for dephasing channel and is achieved for $p=1$ and maximally mixed reduced state of $Q$ with $S_{\rm lin}^Q=0.5$.
	
	\subsubsection{`Both collective' schemes}
	
	As we have already emphasized above, the state $\ket{0}$ is not affected by the amplitude damping channel, 
	that is why we can employ the pre-processing and post-processing collective operators in the form given by Eq.~\eqref{eq:ucoldep} and Eq.~\eqref{eq:vcoldep}, correspondingly.
	As in the case of dephasing channel, they allow achieving the maximal unit fidelity $F_{\rm col\;col}^{\rm opt}=1$.

	\section{Discussion and conclusion}\label{sec:conclusion}
	
	We summarize the main results of the present work. 
	We have suggested the method for improving fidelity based on the class of state-dependent operations protecting the system from decoherence.
	Specifically, we have considered the case where the decoherence affects a single $\kappa$th qubit from an $n$-qubit pure state, while the operators can be employed either on the same $\kappa$th qubit or on the whole state.
	We have shown that two schemes of this class, namely `individual-then-collective' and `collective-then-individual' provide the same maximal achievable levels of the fidelity.
	
	We have considered these schemes for three basic decoherence models given by depolarizing, dephasing, and amplitude damping channels.
	The main results on the comparison between all the strategies of error suppression for various decoherence models are summarized in Table~\ref{tbl}.
	For all the channels we have seen that for given decoherence strength the maximal fidelities in all the schemes except `both collective' is expressed with the linear entropy of the qubit under decoherence $S_{\rm lin}^Q$.
	In particular, we have obtained that the larger is the linear entropy the lower is the fidelity.
	
	This feature provides an answer to the question which qubit from the whole $n$-qubit system is the most vulnerable and which qubit is the most robust in the sense of preserving the whole $n$-qubit state after qubit decoherence.
	It turns out that the best choice for $\kappa$ is defined by the qubit whose reduced state has the lowest linear entropy.
	For the qubit case, the reduced state with the lowest linear entropy is automatically the state with the lowest min-entropy, von Neumann entropy, entropy of entanglement and quantum discord.
	A promising direction for future research may be also to consider other various measures of decoherence~\cite{Fedichkin20032,Fedichkin2004,Fedichkin2009}.

	The obtained results surely have a clear intuitive explanation, since the more the entropy of the reduced state of the qubit is, the more it is entangled to remaining qubits of the system, and the more it is `informationally important' to the whole state.
	However, we would like to mention that one should be very careful with such kind of intuitive conclusions. 
	E.g. if we turn to speak about quantum correlations (say, entanglement), as it was shown in Refs.~\cite{Zyczkowski2001,Kiktenko2012,Korotaev2012}, 
	in the case of depolarizing channel and mixed two-qubit state, the state with the largest entropy may be more robust in the sense of preserving initial correlations.
	
	Finally, we would like to mention that the considered schemes seem to be effective in the framework of quantum communication protocols, where several parties employ an entangled state distributed between them.
	If the communication channels have a different degree of decoherence, then the whole protocol can be adjusted in such a way that the most robust parts of the entangled state go through the noisiest communication paths.
	
	\section*{Acknowledgments}
	
	The work was supported by the grant of the President of the Russian Federation (project MK-923.2019.2).

\end{document}